\documentclass[a4paper,fleqn]{cas-sc}

\usepackage[numbers]{natbib}

\newcommand{\hidecomment}[1]{}
\newcommand{\system}{\textsc{Urban~Chronicles}\xspace}
\newcommand{\myparagraph}[1]{\noindent \textbf{#1}\xspace}
\newcommand{\maryam}[1]{{\color{purple} Maryam: [{#1}]}}

\def\tsc#1{\csdef{#1}{\textsc{\lowercase{#1}}\xspace}}
\tsc{WGM}
\tsc{QE}
\tsc{EP}
\tsc{PMS}
\tsc{BEC}
\tsc{DE}

\begin{document}
\let\WriteBookmarks\relax
\def\floatpagepagefraction{1}
\def\textpagefraction{.001}
\shorttitle{A Visual Analytics System for Profiling Urban Land Use Evolution}
\shortauthors{C. Santos et~al.}

\title [mode = title]{A Visual Analytics System for Profiling Urban Land Use Evolution}                      



\author[1]{Claudio Santos}[orcid=0000-0001-8218-754X]
\ead{c.souza0804@gmail.com}
\author[2,3]{Maryam Hosseini}[orcid=0000-0001-8329-4638]
\ead{maryam.hosseini@nyu.edu}
\author[2]{João Rulff}[orcid=0000-0003-3341-7059]
\ead{jlrulff@nyu.edu}
\author[4]{Nivan Ferreira}[orcid=0000-0001-6631-4609]
\ead{nivan@cin.ufpe.br}
\author[5]{Luc Wilson}
\ead{lwilson@kpf.com}
\author[6]{Fabio Miranda}[orcid=0000-0001-8612-5805]
\ead{fabiom@uic.edu}
\author[2]{Claudio Silva}[orcid=0000-0003-2452-2295]
\ead{csilva@nyu.edu}
\author[1]{Marcos Lage}[orcid=0000-0003-3868-8886]
\ead{mlage@ic.uff.br}



\address[1]{Department of Computer Science, Universidade Federal Fluminense, Niterói, RJ, Brazil}
\address[2]{Department of Computer Science and Engineering, New York University, New York, NY, United States}
\address[3]{Urban Systems, Rutgers University, Newark, NJ, United States}
\address[4]{Centro de Informatica, Universidade Federal de Pernambuco, Recife, PE, Brazil}
\address[5]{Kohn Pedersen Fox, New York, NY, United States}
\address[6]{Department of Computer Science, University of Illinois at Chicago, Chicago, IL, United States}








\begin{abstract}
The growth of cities calls for regulations on how urban space is used and zoning resolutions define how and for what purpose each piece of land is going to be used.
Tracking land use and zoning evolution can reveal a wealth of information about
urban development.
For that matter, cities have been releasing data sets describing the historical evolution of both the shape and the attributes of land units.
The complex nature of zoning code and land-use data, however, makes the analysis of such data quite challenging and often time-consuming.
We address these challenges by introducing \system{}, an open-source web-based visual analytics system that enables interactive exploration of changes in land use patterns. Using New York City's \emph{Primary Land Use Tax Lot Output (PLUTO)} as an example, we show the capabilities of the system by exploring the data over several years at different scales.
\system{} supports on-the-fly aggregation and filtering operations by using a tree-based data structure that leverages the hierarchical nature of the data set to index the shape and attributes of geographical regions that change over time.
We demonstrate the utility of our system through a set of case studies that analyze the impact of Hurricane Sandy on land use attributes, as well as the effects of proposed rezoning plans in Downtown Brooklyn.
\end{abstract}



\begin{keywords}
visual analytics \sep big data \sep urban planning \sep GIS
\end{keywords}

\maketitle

\section{Introduction}
Cities have long been at the center of innovation and development. As people concentrate on these hubs of culture, commerce, and services, hamlets give way to towns and towns to metropolis. This growth, organized or not, calls for regulations and plans to best divide and utilize a precious resource: land. The wheel of time delivers new patterns: carriages evolved to elevated trains reshaping the forms of the cities \citep{huang1996land}; personal cars emerged, redefining the use of urban space by dividing streets between pedestrians and cars \citep{schaeffer1980access}; cabins transformed into skyscrapers, creating the necessity to regulate the size and forms of the buildings and the creation of air rights \citep{black2003urban, marcus1983air}. The modern settlements were faced with a big challenge to adequately manage lands, the very foundation on which cities were built. 
Since then, cities have been adopting and modifying regulations that govern land use and development. Zoning resolutions were defined with the goal of facilitating the management of urban spaces and prescribing specific sets of rules for different pieces of land in the city. They control the shape and size of the buildings and public spaces, preserve urban fabrics, and determine how and for what purpose each piece of land can be used \citep{bassett1940zoning}. The zoning designation can change to comply with the ever-growing demands of urban life and modern cities. 

Tracking the changes in the zoning codes and land use can help drawing a more precise picture of how urban economic, social, and public policies have evolved. It also provides a way to evaluate the performance of the enacted plans \citep{MUNNEKE2005455}. Moreover, the analysis of the zoning evolution at different levels can inform stakeholders of the delayed impacts of some policy changes and major events such as natural disasters or economic crises, market trends, and development opportunities. 
%
%
%
The complex nature of zoning codes, land use regulations, and taxing system makes 
the analyses of such data quite challenging and often time-consuming. For instance, since such extensive data spans different city agencies, each having their own data collection and recording methods, the unified data set is highly prone to errors and inconsistencies.
On top of that, land use and zoning data are often large and complex. In New~York~City~(NYC), the data set describing primary land use is more than 18~GB, has over 83 attributes and more than 320,000 geographical units only for the boroughs of Manhattan and Brooklyn. This data can be considered a spatiotemporal data set, but one where the spatial attribute is not merely composed of points but of complex geometric primitives.
Moreover, new geographical units are created and merged over time, resulting in attributes and geometries change. Tracking these changes is, therefore essential. 
The majority of the existing land use and zoning visualization tools and systems are either merely exploratory tools offering very limited or no analysis capabilities \citep{zola}, or do not take the temporal aspect of the data into account~\citep{masnyc}.

To address some of the challenges involved in the analysis of land use data, we introduce \system, a new visual analytics system for profiling city land use evolution. \system is developed through ongoing collaborations with domain experts (urban planners and architects) that work with land use and zoning data as an integral part of their projects. 
The system contains a data processing pipeline to remove inconsistencies and redundancies common to land use data; an in-memory tree-based data structure specifically designed to handle the geometries common to geographical regions of a city and to drive interactive visualization; and a visual interface that enables the visualization of land use data at different scales, from the land unit level to neighborhood level, giving the user analytical capabilities that would not be possible in a single scale analysis.
Our system makes use of the Primary Land Use Tax Lot Output (PLUTO) data, a recently released spatiotemporal data set that describes the land use information in NYC, for over 15 years. We use it as our running example to introduce the system, although \system can also be used to explore other parcel-based land use inventories.
We demonstrate the effectiveness of \system through detailed case studies set in Manhattan and Brooklyn, two boroughs of NYC. The case studies analyze the impact of the National Flood Insurance Program (NFIP) of 2013, a year after Hurricane Sandy, on residential development activities in lower-income neighborhoods, as well as evaluating the Downtown Brooklyn rezoning plan that aimed at absorbing the office space demand generated by Financial District. 
To summarize, our contributions are:
\begin{itemize}
    \item We introduce \system\footnotetext[1]{The open-source system can be found at \url{https://github.com/Prograf-UFF/urban-chronicles}}, an open-source web-based system that enables urban planners and architects to visually explore large data sets composed of geographical, political, and administrative land unit information.
    \item We propose a data pipeline to automatically extract, clean and process raw land use data.
    \item We adapt a general purpose tree-based data structure to leverage the hierarchical nature of land use data. We show that this data structure allows \system to perform in interactive time when querying large spatiotemporal land use data sets. We demonstrate its effectiveness through a set of experiments, including comparisons with industry-standard GIS database and tool.
    \item We highlight the usefulness of \system through two detailed case studies set in NYC and performed by urban planners and architects. An accompany video demonstrates the system.
\end{itemize}
%

\section{Related Work}
\label{sec:related}
%
We survey the literature from three categories: land use and zoning analysis, interactive visualization, and visual analytics in the context of cities.

\myparagraph{Land use and zoning analysis.}
Recently, an increasing number of new detailed data sets describing taxation, land use, and zoning information has become available by multiple city agencies~\citep{pluto,cmap,berlin} providing new research horizons across various different fields. 
Land use and land cover change have long been studied by scientists from multiple different perspectives. Environmental scientists looked at how such changes can impact ecosystem services~\citep{lawler2014projected}, storm run-off~\citep{bronstert2002effects}, climate change~\citep{kalnay2003impact}, among others~\citep{KREMER2013218,CHEN201753}. In urban planning, land use change can signal gentrification~\citep{lim2013urban, hamnett2007loft}, or threaten the urban fabric of a neighborhood and its historic landmarks \citep{goldman2017preserving}. Zoning change can also impact various urban indicators such as urban growth \citep{goldberg2015game}, affordable housing residential development \citep{getz2017examining} and land values~\citep{MA2020104537}. Hence, predicting land use change has for long been a topic of interest~\citep{du2010case}. 
Land use and zoning data sets, such as PLUTO~\citep{pluto_doc}, Chicago's Land Use Inventory~\citep{cmap}, and Chicago's Zoning District~\citep{chicagoZoning} were used in several reviewed studies and published tools but most of them are limited in their exploratory analysis by using a handful of attributes, short time periods, and coarse spatial granularity~\citep{urbanreviewer,nycommons,secondcity, witlox2005expert}. 
In this work, we propose a visual analytics system that enables the interactive exploration of large temporal land use and zoning data sets in its entirety.

\hidecomment{Land conservation and historic preservation strategies are aimed at protecting these vulnerable sites against the development forces. However, such protection comes with costs since limiting the supply of housing can push the prices up and impact the adjacent neighborhood by transferring the demands and development waves~\citep{newburn2005economics, mccabe2016does, been2016preserving}.}
\myparagraph{Interactive visualization.}
To be effective, visual analytics systems should strive for \emph{interactivity}, with response times below 500ms~\citep{liu2014effects}. Thus, several techniques and data structures have been proposed in recent years to handle interactive queries over large data sets and minimize the latency between user input and visual output. Examples of such proposals include ones specifically designed to leverage the processing power of GPUs~\citep{liu2013immens,zacharatou2017gpu}, optimize I/O and network usage~\citep{stash}, use deep learning for learned indices~\citep{10.1145/3183713.3196909,wang2018neuralcubes}, and data cube models~\citep{lins2013nanocubes, 7858782}. These past works, while handling simpler forms of data points, do not take into account more complex types, such as geometric shapes. We propose a novel approach specifically designed for arbitrary shapes, targeted at the interactive queries of land units that uses less memory and are faster than traditional databases.
%

\myparagraph{Urban visual analytics.}
In the past decade, various tools and systems have been proposed to analyze and explore urban data. These systems are often motivated by open data initiatives requiring city agencies to make their data publicly available through open data portals~\citep{nycopendata,sfoopendata,chicagoopendata,opendata@book2013,doi:10.1089/big.2014.0020}, which played a central role in helping experts to gain a deeper understanding of cities, evaluate policies, and plan developments~\citep{zheng2016visual}. 
Examples of such systems include ones designed to explore human mobility~\citep{Andrienko2013a, ferreira2013visual, andrienko2019visual, sobral2020ontology}, public policy~\citep{conejero2021towards}, air pollution~\citep{deng2019airvis,halkos2019understanding}, urban traffic~\citep{gong2020expert}, public transportation~\citep{palomo2016visually,yu2015iviztrans}, real-estate ownership~\citep{hoang2014towards}, land use labelling~\citep{terroso2020land}, urban change~\cite{10.1145/3313831.3376399}, and shadow impact on public spaces~\citep{8283638}.

Several tools have also been proposed to handle more complex urban data, such as city objects with an associated geometry, like buildings. Their goal is to inform the decision-making process in the urban domain and are being used by researchers, architects, urban planners, developers, and the general public. Examples of such tools include PlaceILive~\citep{placeilive}, designed for the assessment of neighborhood quality, Urbane~\citep{ferreira2015urbane,Doraiswamy:2018:IVE:3183713.3193559} and Vis-A-Ware~\citep{Vis-A-Ware}, to analyze potential development sites, and ArcGIS~\citep{johnston2001using}, a general GIS system. These systems, however, were developed as general-purpose tools to cover the whole spectrum of geographical-based analysis. For more focused tasks, such as land use analysis, these systems do not provide the capability to visually and interactively explore and track the temporal variation of the data sets.

To the best of our knowledge, \system is the first visual analytics system designed to interactively explore spatiotemporal land use data sets, motivated by the real-world needs of urban planners and architects.
\begin{figure}[!t]
    \centering
    \includegraphics[width=1.0\linewidth]{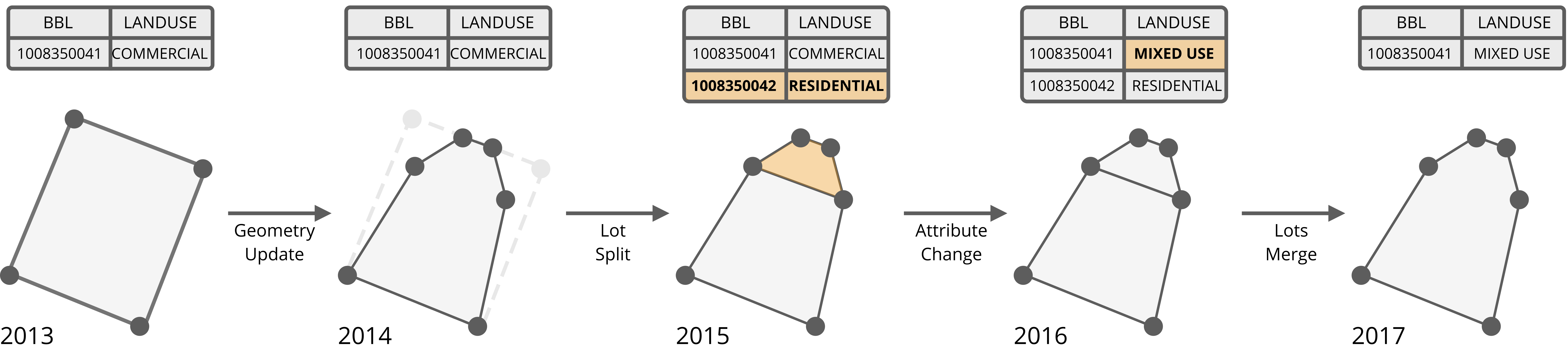}
    \caption{
    Lot changes in PLUTO: The geometry of the lot was updated to better represent its real geometry (2014); In the following year it was split, creating a new lot (2015); In 2016, the larger lot became a mixed-use area; and in 2017 the two lots were merged.  
    }
    \label{fig:geometry-inconsistency}
\end{figure}

\section{Background}
\label{sec:background}
%
The rules governing land use and building size, form, and function are called zoning laws. Zoning puts parcels of land in different groups (or \emph{``zones"}) and prescribes certain sets of regulations for each one of them \cite{morris1989zoning}. 
%
%
There are multiple standards to locate and identify each piece of land. The system used in the U.S. and Canada is called \emph{Lot and Block Survey System}; under which, blocks are defined as \emph{"A tract of land bounded on all sides by streets, or a combination of streets, public parks, railroad rights-of-way, pierhead lines or airport boundaries"}. Blocks are then subdivided into lots, the smallest parcel of land that can be owned, purchased, and sold. In NYC (the running example used in this paper), for property tax purposes, lots are identified with a unique code resulting from the concatenation of borough, block and lot number corresponding to their location, called BBL code. A lot identified with that number is called a tax lot. 
See Figure~\ref{fig:geometry-inconsistency}~(left) for an example of a lot. A lot can be divided to create new lots (2014-2015), or multiple lots can merge to create a larger lot (2016-2017).
\hidecomment{
Two or more adjacent lots within a block can merge and create a larger lot, and a larger lot can also be divided to create new lots, Figure~\ref{fig:geometry-inconsistency}~(middle). \maryam{this does not describe the process we see in Figure 2, it is more matched to what happens in Figure 7}The resulting lot may be assigned a new BBL or carry the BBL of a merged lot~\cite{zoning2018}.
}

%
The zoning designation can change through a process called \emph{rezoning}.
%
The rezoning often seeks to either increase the density and encourage more development by setting \emph{less restrictive} rules on the height, size, design, and use of the buildings; or to prevent density increase, protect local businesses and historic fabric of the neighborhood by setting \emph{more restrictive} zoning rules.
%
%
Tracking the evolution of zoning and how they have influenced different aspects of urban life can provide a wealth of information vital for crafting policies and planning better cities.

\section{Data Description and System Requirements}
\label{sec:data-req}
PLUTO, the NYC's data set that is our motivating example, is a spatiotemporal data whose spatial component of the elements are geographical regions, where each region is described by its polygonal boundary and a list of attributes that change over time. Next, we describe the main features of PLUTO.

\subsection{Primary Land Use Tax Lot Output Data}
PLUTO is an invaluable and extensive land use and geographic data set, describing almost every piece of land in NYC. Provided by the Department of City Planning, the data set contains 83 attributes pertaining to tax lot characteristics, building features, as well as geographic, political, and administrative districts. These attributes are collected and maintained by several city agencies~\cite{pluto_doc}.

MapPLUTO is derived from merging the PLUTO attributes with the Tax Lot Polygon features from the Department of Finance. The data provides a wealth of information about land use and zoning, for instance, land use codes, zoning districts, and maximum allowable floor-to-area ratio (FAR); property tax such as total assessed value, assessed  land value, and tax exemptions; primary building features like building class, frontage and area, among others. It also assigns each lot to a Community District (CD) i.e., administrative districts unique to New York City. 
Both PLUTO and MapPLUTO are updated twice a year and are considered among the most prominent data sets publicly available in NYC.
%
%
%
%
This work is based on the MapPLUTO data set but, for simplicity, we will refer to it as PLUTO data set in the text.

\subsection{PLUTO Data Characteristics}
We collected PLUTO data from 2002 to the first semester of 2017. During this period, 22 versions of the data set were released. The total amount of space required to store all the 22 releases is approximately 18 GB.
Each PLUTO data release is a collection of shapefiles, one for each of the five boroughs in NYC, containing the geometries and attributes of the lots.

Although the raw historical data is large, it contains many redundancies that can be explored for storage optimization. For example, the geometry of most of the lots of NYC remains unchanged for several years. We observed that on average, only 8\% of the lot geometries change in two consecutive releases of the data set. Similarly, the average amount of changes in the lot attributes on subsequent versions is 40\%. 
Figure~\ref{fig:geometry-inconsistency} illustrates all changes that the geometry and attributes of a lot can undergo. We stress that tracking and visualizing the changes of all lots of a city, is a challenging task that we aim to solve with \system{}.
%


\myparagraph{Inconsistencies.}
Like many public data sets, PLUTO contains multiple inconsistencies between different versions of the data making longitudinal analysis cumbersome and time-consuming.
For example, there were cases that the geometry of a given lot did not actually change, while its geometric attributes were updated in the data due to data entry errors or changes in the data acquisition method (Figure~\ref{fig:geometry-inconsistency}~(2013-2014)). In some cases, the overlap between two versions of the geometry of the same lot was smaller than $50\%$.
%
%
Moreover, the attributes of the lots have gone through several changes over the years. Examples of issues one observes when trying to consolidate the different PLUTO releases are changes in attributes names, changes in attribute definition and inaccurate attribute information.  
%
%
%
In Section~\ref{sec:pipeline} we describe our preprocessing strategy to alleviate these issues.
\subsection{System Requirements}
\label{sec:requirements}
In our ongoing collaboration with domain experts from urban planning and architecture, we identified a set of tasks that they are interested in performing:
1)~Study the temporal patterns of change in certain attributes, such as residential and commercial areas or assessed value at the neighborhood, block, or lot level.
This can help detecting anomalies and major events. 
2)~Evaluate the implementation of rezoning plans, whether and when the zoning has changed, if it has been applied to the blocks initially targeted in the plan, and its impact on adjacent neighborhoods.
3)~Identify regions based on certain attributes such as the level of unused residential and commercial FAR, specific land use (e.g., vacant lands), zoning district, building classes,  and compare lot level data in a selected neighborhood or community district. Based on the tasks mentioned above, we defined the following system requirements:


\myparagraph{[R1]~Enable the exploration of data at different city scales.}
Enable experts to explore land use data at different geographical levels (e.g., borough, neighborhood, community district, block, lot). An urban planner might, for instance, use the neighborhood level to identify areas undergoing a process of gentrification and then analyze the phenomena at the block or lot level.

\myparagraph{[R2]~Support the exploration of attribute changes over time.}
Allow users to explore the rate of change in the attributes over time, while enabling the identification of outliers or temporal patterns.

\myparagraph{[R3]~Support the exploration of lots of interest.} 
Being able to filter lots based on tax, land use or zoning attributes allows experts to identify areas that follow certain criteria, such as unused FAR or rezoning status.

\myparagraph{[R4]~Support the exploration of temporal changes of lots.}
Throughout the years, lots are divided or merged. Keeping track of changes in the number of lots on neighborhood, community district, or block level can be of special interest to urban planners. For example, the increase of lot merges can be a signal of an incentive zoning program or transferable development rights, which can have a large impact on the community.
%

\myparagraph{[R5]~Support interactive query times.}
All the previous tasks should be executed interactively since response time greater than 500ms can significantly impact visual analysis, reducing the rate at which users make observations~\citep{liu2014effects}.
%
%
\begin{table}[t]
    \centering
    \caption{The PLUTO data set attributes selected by our partners urban planners and architects to be explored using our system.}
    \vspace{0.1cm}
    \scalebox{0.85}{
    \begin{tabular}{|l|l|l|}
        \hline
        Attribute Class  & Variations                  & Description \\
        \hline
        \hline
        BBL              &  --                         & The concatenation of the borough, block and lot codes.\\
        \hline
        AREA             &  Lot, Residential, etc.     & The area of the lot, the area allocated for residential use, and other area measurements. \\
        \hline
        ASSESS           &  Land, (Normalized) Total   & The assessed land and total value for the lot. \\
        \hline
        BLDGCLASS        &  --                         & A code describing the major use of structures on the lot. \\
        \hline
        FAR              &  Built, Residential, etc.   & The built or maximum allowed floor area divided by the area of the lot.\\
        \hline
        LANDUSE          &  --                         & A code for the tax lot's land use category. \\
        \hline
        NUM              &  Buildings, Floors          & Number of buildings and number of floors in the tallest building of the lot.\\
        \hline
        SPDIST           &  --                         & The special purpose district which can impact the underlying zoning codes. \\
        \hline
     \end{tabular}
     }
    \label{tab:attributes}
\end{table}
\begin{table}[th]
    \centering
    \caption{Percentage of redundant lot geometries and attribute sets in the PLUTO releases and memory usage. The amount of  redundant lot geometries is higher than 90\%. The overhead of the \emph{Data Storage} used by \system{} is, in the worse case, 3.5 times the amount of space required to store the data produced by the \emph{Preprocessing Pipeline}. To evaluate the storage efficiency of \system{} we loaded the prepossessed data into a relational database. The space required by the database is, in the best case, 5.11 times the data size.}
    \vspace{0.1cm}
    \scalebox{0.85}{
    \begin{tabular}{ |c|c|c|c|c||c|c|c|c||c|c| }
    \cline{2-11}
    \multicolumn{1}{c|}{} & \multicolumn{4}{c||}{Redundancy (\%)}&\multicolumn{4}{c||}{Memory Usage (GB)} & \multicolumn{2}{c|}{Overhead} \\
    \cline{2-11}
    \multicolumn{1}{c|}{} & Geometry & Unstable & Stable & Categorical & Raw & Pre (P) & Storage (S) & DB (D) & S/P & D/P \\
    \cline{1-11}
    Manhattan & \textbf{91}  &      28     &       75      &      \textbf{79}      & 0.53 & 0.28 & 0.93 & 1.43 &      3.33    & \textbf{5.11}\\
    \hline
    Brooklyn  &      92      &      \textbf{21}     & \textbf{66}   &      \textbf{79}      & 3.28 & 1.86 & 6.50 & 10.0 & \textbf{3.50} & 5.37\\
    \hline
    \end{tabular}
    }
    \label{tab:memory}
\end{table}
\section{\system{} System}
\label{sec:system}
In this paper we present the visual analytics system called \system{}, developed to enable the exploration of land use data. The system is composed of three modules: the \emph{Data Storage}~(Section~\ref{sec:storage}), the \emph{Query Processor}~(Section~\ref{sec:query}) and the \emph{Visual Interface} ~(Section~\ref{sec:visual}).
%
%
Figure~\ref{fig:scales} highlights the different components of the visual interface. An accompany video demonstrates the main features of the system.
Before describing the system in detail, we present the \emph{Data Preprocessing Pipeline}, a set of steps that prepares the PLUTO data to be consumed.

\subsection{Data Preprocessing Pipeline}
\label{sec:pipeline}
\begin{figure}[t!]
    \centering
    \includegraphics[scale=0.08]{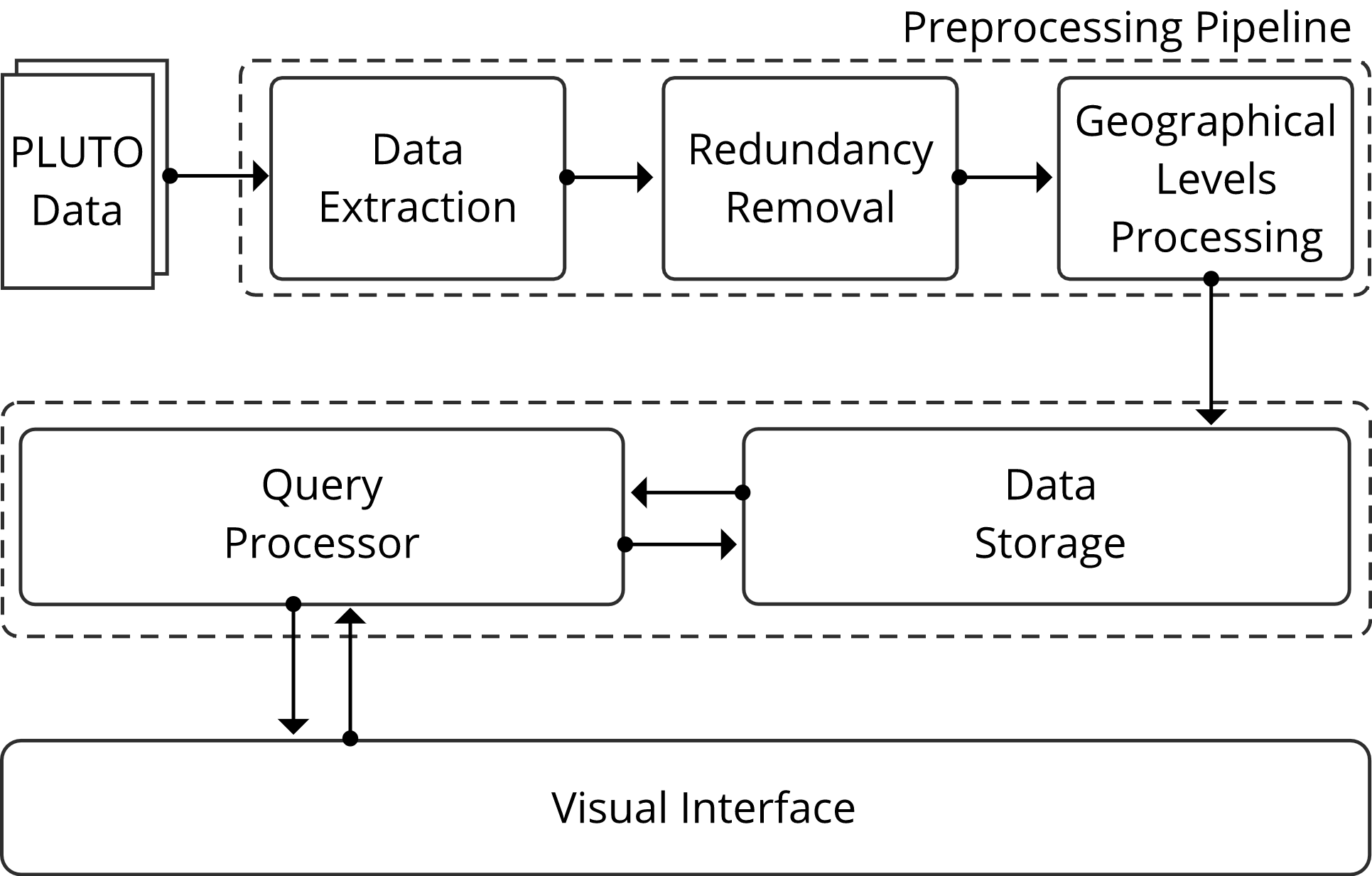}
    \caption{
    The \system{} modules. The \emph{Preprocessing Pipeline} prepares the raw PLUTO data to be stored by the \emph{Data Storage}. As long as the user interacts with the \emph{Visual Interface} the produced queries are handled by the \emph{Query Processor}.}
    \label{fig:system-overview}
\end{figure}

%
Due to the inconsistencies described in Section~\ref{sec:data-req}, a preprocessing phase is needed to standardize the data and enable its consumption by our system.
To achieve this goal, the \emph{Data Preprocessing Pipeline} implements the three steps indicated in Figure~\ref{fig:system-overview}: data extraction, redundancy removal, and geographic level processing. 

\myparagraph{Data extraction.} This step of the pipeline is responsible for collecting, cleaning, and consolidating the information contained in the raw collection of shapefiles. 
%
The cleaning task consists of discarding invalid lots (e.g., lots with missing geometric components) from the data set and standardizing attribute names that changed over the years.

The collecting task is responsible for extracting from PLUTO the attributes that will be made available by \system{} for exploration. The attributes used in this paper were selected by our urban planners and architects collaborators and are shown in Table~\ref{tab:attributes}.
Finally, the consolidation task is responsible for merging the collected and cleaned data of all boroughs into a single CSV file that is the input of the next step of the pipeline.

\myparagraph{Redundancy removal.} The goal of this step is to detect and discard redundant geometries and attributes contained in the releases of PLUTO. 
We observe that the shape of most lots doesn't change over the years. When that happens, we store the geometry of the first occurrence of the lot and replace the geometry of the following occurrences with a reference to the stored one.
However, as discussed in Section~\ref{sec:data-req}, the geometry of the lots in the PLUTO data set may be updated even when they remain unchanged in the real world, revealing a data quality problem. To handle this issue, we use the following strategy:
Let's say that $\mathtt{b}_i$ denotes a lot with BBL $\mathtt{b}$ in a given year $\mathtt{Y}_i$ and $\mathtt{g}(\mathtt{b}_i)$ denotes the geometry of $\mathtt{b}_i$, where $i \in [1, |\mathtt{Y}|]$, $\mathtt{Y} = \{2002,...,2017.1\}$. We say that $\mathtt{g}(\mathtt{b}_{i+1})$ and $\mathtt{g}(\mathtt{b}_i)$ are the same when:
$$\frac{\mathtt{Area}(\;\mathtt{g}(\mathtt{b}_i) \cap \mathtt{g}(\mathtt{b}_{i+1})\;)}
{\mathtt{max}(\;\mathtt{Area}(\;\mathtt{g}(\mathtt{b}_i)\;),\; \mathtt{Area}(\;\mathtt{g}(\mathtt{b}_{i+1})\;)\;)} \leq \epsilon$$  
In our implementation we used $\epsilon = 0.9$. As we show in Table~\ref{tab:memory}, the average geometry redundancy on each borough is higher than 90\%.

A similar idea is applied to remove attribute redundancy. However, instead of individually handling each attribute, we group them into sets of attributes and track unchanged sets. 
We define three types of attribute sets called categorical, numerical-stable, and numerical-unstable.
As the name suggests, the categorical set includes the categorical attributes of the lots, such as the \emph{LANDUSE} class. 
The numerical-stable set includes numerical attributes that change less frequently over time, such as the \emph{LOTAREA} value. 
The numerical-unstable set includes numerical attributes that are likely to vary over the years, such as the \emph{ASSESSLAND} value. 
Lets denote by $\mathtt{a}(\mathtt{b}_i)$ the values in a set of attributes of a lot with BBL $\mathtt{b}$ in year $\mathtt{Y}_i$. 
If all values of $\mathtt{a}(\mathtt{b}_i)$ and $\mathtt{a}(\mathtt{b}_{i+1})$ are the same, we store the set $\mathtt{a}(\mathtt{b}_i)$, and represent $\mathtt{a}(\mathtt{b}_{i+1})$ by a reference to $\mathtt{a}(\mathtt{b}_i)$. 
On average, at least 79\% of the categorical, 66\% of the numerical-stable and 21\% of the numerical-unstable sets are redundant (see Table~\ref{tab:memory}).

\myparagraph{Geographic level processing.}  
Several different geographical subdivisions are used by the public administration in the definition of city policies, rezoning plans, etc.
%
%
For this reason, \system{} provide to users the ability to explore PLUTO data based on the borough, neighborhood, community district, block, and lot levels (see requisite \textbf{R1}).  
This requisite requires the computation of the lots that belong to the geographical regions of each level. 
PLUTO data already stores to which borough and block belong every lot of the city (see Section~\ref{sec:background}). 
%
%
On the other hand, the neighborhood and community district that contains a given lot must be computed using geometric operations. 
More precisely, given a geographical subdivision $\mathtt{P}=\{\mathtt{p}_1, ..., \mathtt{p}_{|\mathtt{P}|}\}$, we compute pairwise intersections between the polygonal regions $\mathtt{p}_i$ and the blocks of the city on every year.
To do so, we randomly select a lot $\mathtt{b}_j$ and test its intersection with all regions $\mathtt{p}_i$.
Once we find a region $\mathtt{p}_i$ that contains the lot $\mathtt{b}_j$, we say that all lots in the same block of $\mathtt{b}_j$ also belong to $\mathtt{p}_i$.
%
\subsection{Data Storage}
\label{sec:storage}
\begin{figure}[t!]
    \centering
    \includegraphics[scale = 0.18]{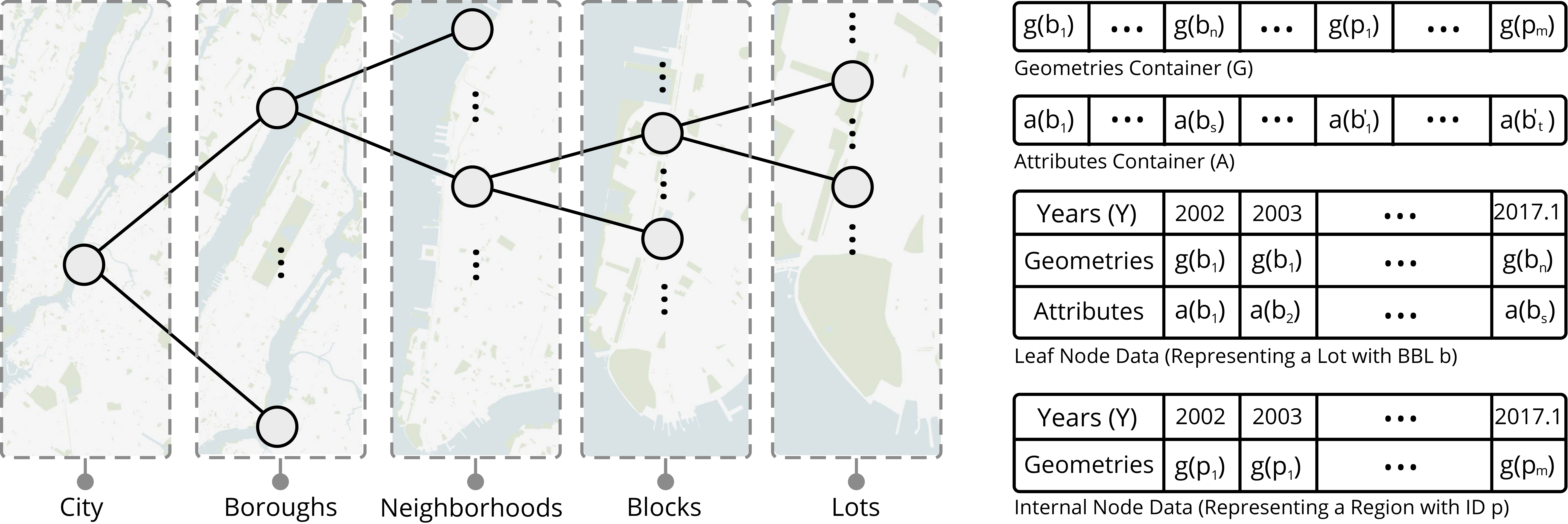}
    \caption{The data transformed by the \emph{Preprocessing Pipeline} is stored in a custom data structure composed by the \emph{Spatio-temporal Index}, the \emph{Geometries Container} and the \emph{Attributes Container}.
    The first uses a tree data structure to locate the lots at each geographical level. The second and third store the geometries and attribute sets extracted from the PLUTO data set releases. 
    The internal and leaf nodes of the tree, store references to the data associated with the geographical regions and lots that they represent and are stored in containers. 
    }
    \label{fig:data_structure}
\end{figure}

%
After preprocessing the PLUTO data set, the transformed information is stored in a custom data structure, illustrated in Figure~\ref{fig:data_structure}.
The data structure, composed of in-memory \textit{data containers} and a \textit{spatiotemporal index}, was designed to work together with the \emph{Query Processor} module to handle the requests generated by the \emph{Visual Interface} module (see Section~\ref{sec:visual}). This data structure allows our system to process queries in interactive time, allowing us to tackle one of the requisites \textbf{[R5]} listed in Section \ref{sec:data-req}.

%


%
The \textit{data containers} store the lots' and geographical regions' geometries, as well as the lots' attributes sets that were extracted in the \emph{Preprocessing Pipeline}.
We call these data containers as \emph{Geometries Container} and \emph{Attributes Container}. 
%
The \emph{Spatiotemporal Index} stores the geographical level information computed in the \emph{Preprocessing Pipeline}.
It is implemented using a tree data structure composed of five levels that represent the city, boroughs, neighborhoods / districts, blocks, and lots' geographical scales.
We observe that either the neighborhoods or the community district regions are taken into account during the construction of the data structure. Since this option has no impact on the structure's design, from now on, we will assume the neighborhood regions are chosen.
We also noticed that a geographical region in level $l+1$ is always contained in a single region of level $l$. This property ensures that the different levels can be properly represented using a tree. 
In this way, the city level is represented by the root node of the tree and covers the entire NYC. Analogously, the \emph{borough} level contains nodes that represent each of the five boroughs of the city; the \emph{neighborhood} level contains nodes that represent the neighborhoods of each borough; and so on.
Each leaf node stores one reference to its \emph{Leaf Node Data}, which contains the indexes of the \emph{Geometries Container} and \emph{Attributes Container} positions in a global list that store the lot's geometry and attribute sets in each PLUTO release.

Similarly, the internal nodes of the tree store one reference to its \emph{Internal Node Data}, which contains the index of the \emph{Geometries List} positions that hold the region's geometry in the year of each PLUTO release.
The amount of memory required by the \emph{Spatiotemporal Index} and the data containers are presented in Table~\ref{tab:memory}. As we observe, the overhead is, in the worse case, 3.5 times the amount of data occupied by the preprocessed data. 

To verify the efficiency of our data structure, we compare it to a widely adopted standard solution to store and query datasets. 
More precisely, we compare both the loading time and memory usage of the preprocessed data using the object-relational database PostgreSQL. 
We defined a relational schema to accommodate our data model in PostgreSQL that comprises of 5 different tables to ensure data normalization. 
We observed that our solution provides both better loading times and storage consumption. 
As an example, for the data subset related to Manhattan, the loading time for all geometries took 17 seconds, while in PostgreSQL it took 626 seconds. 
Similarly, we can also observe an improvement in storage consumption. While our solution only uses 992~MB of in-memory space, PostgreSQL uses 1429~MB of space shared between memory and disk space.

\subsection{Query Processor}
\label{sec:query}
%
The query processor uses the \emph{Data Storage}'s spatiotemporal index and containers to process the queries produced by the visual interface. The queries supported by the system can be classified in geometry retrieval, attribute aggregation, and lot filtering operations.   
The geometry retrieval queries return the shape of a set of geographic regions in a given year. For example, if the user wants to retrieve the geometry of the lots that belong to the block with id 01036 in the Brooklyn neighborhood called Park Slope in 2006, the \emph{Visual Interface} would send a request to the \emph{Query Processor} with the tuple \textsc{(NYC, Brooklyn, Park Slope, 01036, 2006)} as a parameter.
The \emph{Query Processor} then parses the tuple, navigating through the spatiotemporal index to collect the data.

The attribute aggregation queries return an attribute of interest aggregated over a set of geographic regions. For example, if the user wants to see the sum of the LOTAREA for the lots in each block of a Manhattan neighborhood called SoHo in 2009, the \emph{Visual Interface} would send a request to the \emph{Query Processor} with the tuple \textsc{(NYC, Manhattan, SoHo, LOTAREA, SUM, 2009)} as a parameter for the \emph{Query Processor} to return the result.
%
Our system supports distributive (e.g., sum, count, min, max) and algebraic (e.g., average) aggregate functions.

The spatiotemporal index and aggregations are constructed in parallel in real-time, allowing users to define filters on the PLUTO data and then analyze lots that satisfy certain constraints. 
For example, if the users are interested in studying residential lots, they can define a filter using the \emph{Visual Interface} and ask the \emph{Query Processor} to rebuild the index considering only lots that fit the desired criteria. Users can compose filters using boolean operations.
Using a notebook equipped with a Core i7@2.2~GHz, 16~GB of RAM, SSD drive, and NVidia~GeForce GTX~1060~GPU,
the time required to compute summaries and filter lots for an entire borough like Manhattan is on average 0.3s, while in PostgreSQL, using the schema cited in \ref{sec:storage}, it took 4.86s for the same query. We also calculated the time taken for similar queries in neighborhood and block levels. Queries using our method took on average 0.03s and 0.01s, respectively. Using PostgreSQL, the same queries took on average 0.85s and 0.86s. All times reported in this section were calculated based on an average of 10 trials of the same query.
%
\subsection{Visual Interface}
\label{sec:visual}
\begin{figure}[t!]
    \centering
    \includegraphics[width=0.85\linewidth]{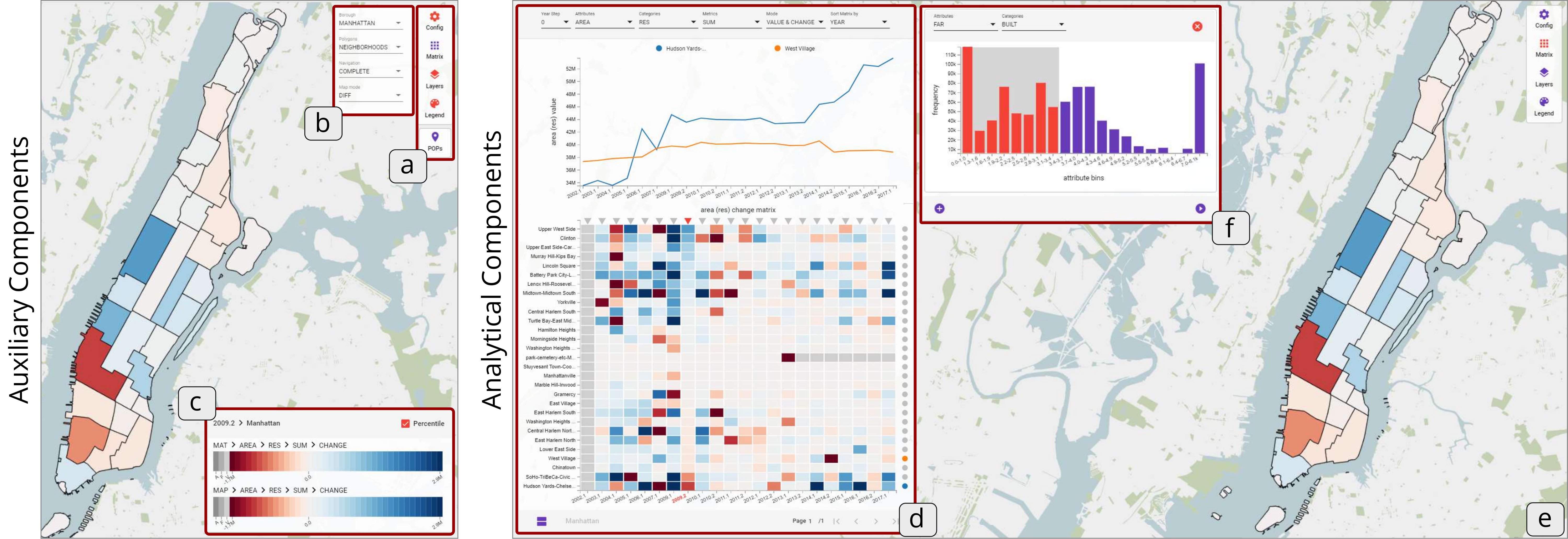}
    \caption{The components of the \system{}' \emph{Visual Interface} can be separated into two groups.
    The image on the left shows the \emph{Auxiliary Components} group, composed by the \emph{Main Menu}~(a), the \emph{Configuration Panel}~(b) and the \emph{Color Maps Legend}~(c).
    The image on the right shows the \emph{Analytical Components} group that contains the \emph{Data History View}~(d), the \emph{Map View}~(e) and the \emph{Lot Filtering}~(f).
    }
    \label{fig:interface}
\end{figure}

%
Based on the requirements specified in collaboration with our urban planners and architects collaborators (Section~\ref{sec:data-req}) we designed a visual interface that allows users to interactively explore the history of PLUTO data at different spatial scales. 
%
%
%
A coarser geographical level provides the user with an overview of the PLUTO attributes. 
For example, at the neighborhood level, it is possible to identify what neighborhood has the highest increase in the commercial area in a year of interest. 
Moreover, whenever more detail is required, the users can interactively drill down to the block and lot levels. Using the previous example, the user could explore the lots of the neighborhood and check if a new commercial center was built. 
As shown in Figure~\ref{fig:interface}, the visual interface is composed of several components. These components can be separated into two groups, the auxiliary and the analytical ones, described in the following.
%
%

\myparagraph{Auxiliary components.}
%
This group contains the \emph{Main Menu}, the \emph{Configuration Panel} and the \emph{Color Maps Legend}, labeled as (a), (b) and (c) respectively in Figure~\ref{fig:interface}. 
The \emph{Main Menu} allows the user to show/hide the other components. The \emph{Configuration Panel} lets the user adjust the system options. 
More precisely, in this panel, it is possible to define what borough the user wants to explore; what set of geographical regions (neighborhoods or community districts) should be used in the coarser scale of the analysis; whether the block levels should be skipped or not during the analysis; and the color mapping strategy used in the \emph{Map View} (see details below).

\begin{figure}[t]
    \includegraphics[width=\linewidth]{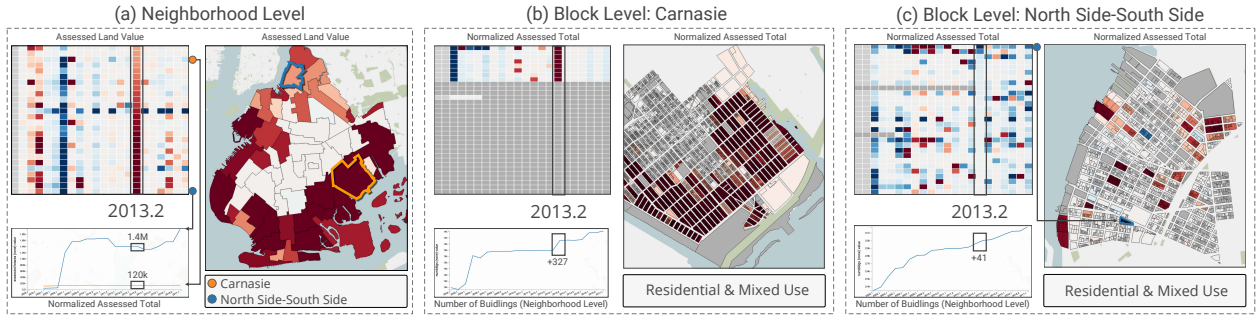}
    \centering
    \caption{Example of interactive exploration of NYC's historical land-use data at different spatial resolutions using \system{}.
    %
    (a) Neighborhood Level: Temporal evolution of the assessed land value of Brooklyn neighborhoods.
    The heat matrix encodes value change in each neighborhood (rows) over the years (columns of the matrix).
    Red cells mark significant drop in assessed land value in the flood zone areas in 2013, a year after Hurricane Sandy.
    The map depicts the spatial variation of the value decrease, where the neighborhoods with the highest drop are on the shoreline.
    (a-bottom left) The line chart show a sharp difference between the normalized assessed total value of the two neighborhoods: Carnasie (Orange line) and North Side-South Side (Blue line). 
    (b,c - top) Block level: North Side-South Side has some blocks with increased normalized assessed total value while in Canarsie, we see severe value drop in most residential blocks. 
    Line charts depicts the new residential buildings constructed in 2013. 327 new buildings were constructed in Canarsie (b-bottom left) while North Side-South Side only had 41 new buildings (c-bottom left).
    %
    }
    \label{fig:scales}
\end{figure}

\myparagraph{Analytical components.}
%
This group contains the \emph{Data History View}, the \emph{Map View} and the \emph{Lot Filtering} components, indicated respectively by the labels (d), (e) and (f) in Figure~\ref{fig:interface}. 
%
The \emph{Data History View} contains two linked visualizations, a line chart and a heat matrix. Used together, they enable the user to explore the value of lot attribute of interest and its change over the years (see requirement \textbf{R2}). The user can select the attribute that will be visualized in both charts using the top toolbar.
Each line of the heat matrix represents a geographical region of the current exploration level. For example, if the neighborhood level is active, each line of the matrix represents one neighborhood of NYC. Also, each column of the matrix is associated with a release year of the PLUTO data.
By default, the lines of the matrix are sorted using the alphabetical order of the names of the regions, but they can also be sorted based on the values of a given year.
In the example of Figure~\ref{fig:interface}, the heat matrix is sorted by the values of 2011.1. 
Using the top toolbar, the color assigned to each square of the heat matrix can be configured to encode the changes of the selected attribute yearly or over a fixed interval of years. Whenever the value of a square is invalid or the lot doesn't exist in the associated year, it is colored in gray. In the example of Figure~\ref{fig:interface}, the first column of the matrix is colored in gray since the \emph{AREA} attribute was not available in 2002.
%
%
The heat matrix allows users to simultaneously analyze the change (or the value) of the selected attribute in all regions/years and facilitates the identification of prominent values. For instance, in Figure~\ref{fig:scales} we observe a major drop in assessed land values of several Brooklyn neighborhoods, blocks and lots in 2013, a year after Hurricane Sandy.

Similarly, each line in the line chart represents a region of the current exploration level.
In Figure~\ref{fig:interface}, the line chart contains two lines representing Lincoln Square (blue) and Hudson Yards (orange) neighborhoods.
The horizontal axis of the chart spans the release dates of the PLUTO data, while the vertical axis can be configured (using the toolbar on the top of the component) to represent the attribute values or their yearly variation.
It should be clear that the line chart and the heat matrix always display complementary information. For example, if the heat matrix is configured to show the attributes variation, the line chart is automatically set to show the attribute values and vice versa. 
The line chart allows the user to easily compare the attribute values (or variation) of regions of interest over the years. The line charts in Figure~\ref{fig:scales} show the variation on the assessed value for selected regions in the neighborhood, block and lot levels. We observe that the value drop can be seen in all scales. 

The \emph{Map View} shows the set of geographical regions of the current exploration level in one of the available years (see requirements \textbf{R3} and \textbf{R4}).
In Figure~\ref{fig:interface} the \emph{Map View} shows the neighborhoods of Manhattan colored using the 2009.2 data.
Similarly to the heat matrix, regions can be colored based on the attribute values or their variation between the selected and the previous release years. In Figure~\ref{fig:interface}, the regions are colored by the attribute variation.
The visualization of geographical regions using the map allows the user to identify spatial patterns that would be impractical to identify using the heat matrix or the line chart. For instance, in Figure~\ref{fig:scales} if we inspect the \emph{Map View} we discover that the drop in the assessed land values happens in the shoreline neighborhoods. On the block and lot levels, it becomes evident that the change happened to blocks and lots located on the peripheral areas closer to the shoreline.

The last analytical component is the \emph{Lot Filtering} panel. This component allows the user to define filters based on the PLUTO data attributes. The filters are used to dynamically select lots that will be considered during the visual analysis. The user is allowed to define multiple filters and combine them using boolean operations. 
It is important to stress that, once a filter is defined, the \emph{Spatiotemporal Index} of the \emph{Data Storage} module is updated in real-time based on lots that satisfy the constraints.
For example, if the users are only interested in analyzing lots with the smallest constructed areas, they can define a filter based on the \emph{BUILTFAR} attribute. As we can see in Figure~\ref{fig:interface}, to defined a filter, a bar chart with the distribution of the values of the selected attribute is shown and the user defines the constraint by brushing over the chart.\\

\myparagraph{Geographical level navigation.}
%
As we discussed before, one of the main characteristics of \system{} is enabling the exploration of the PLUTO data at different scales.
By default the \emph{Visual Interface} module starts in the neighborhood (or community districts level) exploration scale. However, the user can drill down to the block and lot level to explore details in a specific geographical region. 
To do so, the user must click on the label of one of the geographical regions in the vertical axis of the heat matrix.
Once a label is selected, the \emph{Data History View} and the \emph{Map View} are updated to display the data from the selected region. In the example shown in Figure~\ref{fig:scales}, the user started the analysis exploring Brooklyn neighborhoods. After the assessed land value drop was detected, the user clicked in the label of the Canarsie neighborhood. During the inspection of the block level, it became clearer that the drop was concentrated in the shoreline blocks. By clicking in the label of the blocks the user could observe the most affected lots.  
Whenever the user starts to navigate through the different scales, the bottom toolbar of the \emph{Data History View} displays the analysis path that is being followed by the user (see the bottom of component (d) in Figure~\ref{fig:interface}). This way, once the block or lot level is active, the navigation path can be used to roll-up to the coarser levels. 
%
\subsection{Implementation Details}
%
\system was implemented as a web-based system so that we could easily make it accessible to our collaborators.
The back-end of the system was implemented in C++ and consists of a web-server that delivers the web application files and handles HTTP requests by calling the functionalities of the \emph{Data Storage} and \emph{Query Processor} modules.
The \emph{Visual Interface} was implemented using JavaScript.
In order to implement the \emph{Map View} component, we used WebGL and data collected from OpenStreetMap. We decided to implement our custom map solution in order to enable the rendering of a large number of polygons.
The \emph{Data History View} and \emph{Lot Filtering Panel} were implemented using D3, and the \emph{Preprocessing Pipeline} using Python.
\section{Case Studies}
\label{sec:case-studies}
Tracing data back and heeding its trends can tell multiple stories, signal a turning point, or reveal hidden patterns. Besides, we can identify the immediate or deferred impacts of events as well as the trajectory of future trends. All of this can be lost if we only look at cross-sectional data, without gaining any insights into the journey it had through time. 
\system{} enables us to explore intricate data sets to discover these hidden links and build narratives.
In this section, we present two case studies carried out by urban planners and architects that illustrate the utility of the tool. 
\begin{figure}[t!]
    \centering
    \includegraphics[width=1\linewidth]{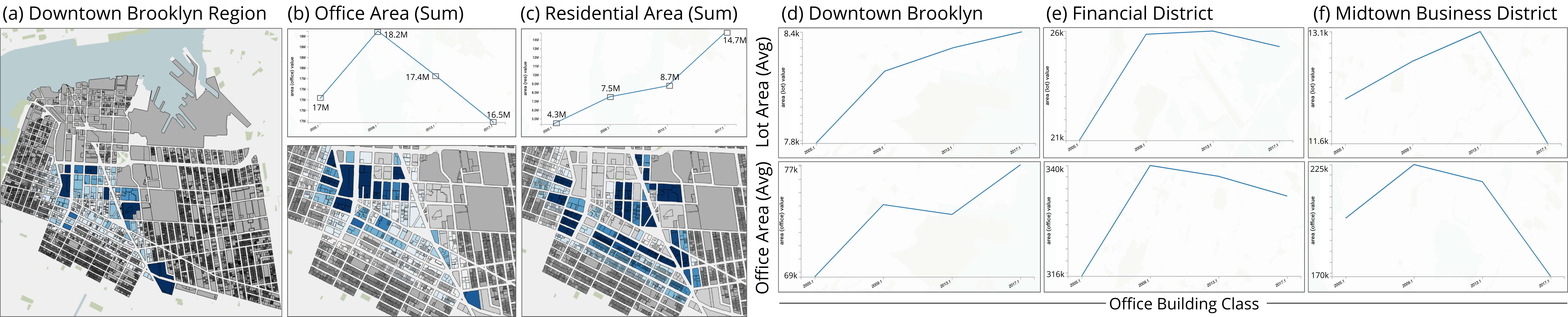}
    \caption{Using \system{} to select lots inside the Special Downtown Brooklyn District (SDBD) (a). Top: Total area of offices (b), and residential (c) buildings in SDBD from 2005 to 2017. Bottom: office (b) and residential lots (c) within SDBD, highlighted in white and shades of blue, darker color indicates larger total area. Line charts depicting average lot area (top) and average office area (bottom) of SDBD, FD and Midtown Business District (d,e,f)}
    \label{fig:case-study}
\end{figure}
\subsection{Post Sandy Redevelopment and Affordability}
Hurricane Sandy hit NYC in late 2012 and the impact was devastating; destroyed buildings, flooded streets, and billions of dollars in damage. In 2013, the National Flood Insurance Program (NFIP) released the Preliminary Flood Insurance Rate Maps (PFIRMs) requiring homeowners in the high-risk zones to pay insurance premiums based on their flood risk. The new program extended the geographical location of high-risk areas. While this can help mitigate the adverse effects in the future, such provisions can create significant pressure for lower-income communities and working-class homeowners living in high-risk flood areas. 

By looking at the data after Hurricane Sandy using \system{}, we can identify patterns that provide insights into the impacts of such provisions on lower-income neighborhoods. We traced the land and property assessed values as well as residential development rates in both areas and discovered alarming trends. 

\myparagraph{Neighborhood level: Assessed land value drop.}
We first found noticeable change of assessed land value in Brooklyn through the \emph{Data History View}. A sharp decrease in the second half of 2013, one year after Sandy, can be easily detected~(Figure~\ref{fig:scales}(a)). As the map shows, this decrease happened in shoreline neighborhoods impacted the most by the hurricane. The pattern also matches the PFIRMs. Next, we use \system{} to explore two of the impacted areas that had significantly different rates of normalized assessed total values while both located in high-risk flood zones. The normalized assessed total value, calculated by dividing the assessed total value by the lot area, gives a more accurate measure when it comes to cross-comparison between regions of different sizes. 
We chose North Side-South Side, a neighborhood with the assessed total value of 1.45 million dollars per sq.~ft and Canarsie, a lower-income neighborhood with 148 thousand dollars assessed total value per sq.~ft. as of 2013. Figure~\ref{fig:scales}(a) shows their location and normalized assessed total values. 

\myparagraph{Block level: Tracing assessed total value change.}
Using \system{}, we drill down to the block level to explore the pattern of value drop in residential areas within both neighborhoods. As shown on the map~(Figure~\ref{fig:scales}(b,c)), while in Canarsie, we see a significant decrease in the normalized total assessed value of the majority of the residential blocks (highlighted in shades of red); in North Side-South Side, some blocks still had their values appreciated (highlighted in shades of blue). One potential explanation can be the larger extent of Sandy damage in Canarsie due to lower flood resiliency measures in residential building construction and lack of coastal protection \citep{canarsie}. This indicates how natural disasters can create more burden for lower-income neighborhoods such as Canarsie, where the majority of the residents are working-class homeowners already struggling with their mortgages \citep{forclosure}. 

Reports show that only five percent of the residential buildings in the whole Canarsie had flood insurance at the time of Sandy \citep{canarsie,risingtides}. The new provision would require all homeowners in the flood zone areas to have flood insurance, and the premium would rise 18\% per year, phasing up to the full premium price. It can be challenging for many households with fixed income to afford the annual 18\% increase and can lead to their displacement. The change in the number of new buildings within the residential areas shows that in 2013, 327 new buildings were constructed in Canarsie~Figure~\ref{fig:scales}(b-bottom left) while in North Side-South Side only 41 new buildings were constructed~Figure~\ref{fig:scales}(c-bottom left). 
The large number of new constructions, specifically the addition of 224 new homes in 2016, the year that PFIRMs took effect, can signal a rise in foreclosure and short sales in Canarsie, possibly due to the inability of homeowners to accommodate the rising mandatory flood insurance together with the repair costs for the majority of uninsured owners for whom the tax decrease is insignificant in comparison to rising tides and costs \citep{risingtides}. Further research can uncover the reasons behind the heightened development activities and examine the role of different factors such as ease of access and proximity to Manhattan or flood resiliency and robust structural system in explaining different rebuilding rates in the two neighborhoods. 

Through analysis such as this one, it becomes possible to evaluate the interplay between different elements such as land value, total assessed value per sq. ft and the number of new buildings. The unique features of \system enable us to interactively switch between geographical levels while tracking all the fine level activities and reveal some unexpected patterns. This leads to insights, and ultimately questions around equitable rebuilding, displacement and affordability that might otherwise not be addressed. 
\subsection{Downtown Brooklyn Rezoning}
In 2004, amidst the fall in economic activities, NYC found its regional position in decline. After 9/11, it was thought that workers would not be willing to return to the Financial District (FD), so the city needed to create a new commercial district to keep jobs from moving to upcoming commercial districts in other areas, specifically New Jersey's new and upcoming neighborhoods in Hoboken and Jersey City. While the prices in Manhattan were still rising and many of the available office buildings did not satisfy the ``Class A'' standards, Brooklyn's plan to transform the Special Downtown Brooklyn District (SDBD) into a vibrant Business District, attracting headquarters from competing places, seemed very plausible.

The goal of the Downtown Brooklyn rezoning was to absorb the office demand that resulted in the decline of the FD and address the NYC's shortage of Class A office inventory by "offering Class A space without Class A rents"~\citep{dtbrezone2004}. 
The plan was approved in 2004, and the final Environmental Impact Statement (EIS) predicted that by 2013 Downtown Brooklyn will 1) construct 4.6 million sq.~ft. of office space, 2) construct 0.9 million sq.~ft. of residential space, 3) increase tax revenue, and 4) increase the public space and cultural amenities.
In a report published by the Brooklyn Borough President in 2016, the outcomes of the rezoning plan after a decade was analyzed, and it was concluded that the plan to a large extent had diverged from its initial goals \citep{Decade2016}. The anticipated addition of 0.9 million sq.~ft. of residential space was met way before the planned time and as the line chart shows~(Figure~\ref{fig:case-study}(c)) and it grew unprecedentedly beyond predictions. This implies that the city should establish regular reevaluations of their goals. 
The rezoned area of Downtown Brooklyn is part of the Community~District~2 (CD2)~(Figure~\ref{fig:case-study}(a)). 
%
\system{} can help discover some of the underlying causes of mismatches between the EIS prediction and the actual outcomes of the plan, looking through the 13 years of land use and tax lot from 2004 to 2017.

\myparagraph{Office and residential space.}
The main motivation of the rezoning plan was to promote the economic viability of the area by providing ample high standard commercial and office space to capture the regional growth and attract private investment. 
As stated in the report, by 2016, only 1.3 million sq.~ft of commercial space (including office space) has been developed in the SDBD \citep{Decade2016}.  PLUTO dataset provides information on the office areas per lot, using \system{} we can take a closer look at the office space development trend. 
The line chart in Figure~\ref{fig:case-study}(b) shows that the total office area in Downtown Brooklyn increased in the first four years of the program but after that, the overall trend has been declining, to the point that in 2017, it declined to less than its initial value in 2004, which is in contrast with the main goals of the plan. 

\hidecomment{As it was mentioned, a major decline in the demand for the FD office spaces was predicted after 9/11. The rezoning aimed at absorbing that demand, but it was not very successful in attaining that goal. }
By comparing the characteristics of the office buildings in FD and Midtown Business District (BD), which absorbed a significant portion of the demand from FD \citep{haughwout2005exogenous}, we can have a better picture of the type of office space in demand. One of the defining characteristics of the FD office space was their large floor plate \citep{furst2006empirical}. \hidecomment{As it was also mentioned in the rezoning plan, the relocating firms were looking for large spaces \citep{dtbrezone2004}.} Looking through office building class, as Figure~\ref{fig:case-study}(e) shows, in FD office buildings were constructed on lots that are on average 25,000 sq.~ft., whereas this number is around 8,000 for SDBD (d-top), more than three times lower than the average lot sizes that FD offices needed. Although Midtown BD also have its offices built on smaller lots of 12,500 sq.~ft. on average (e-top), it is still significantly higher than SDBD. Moreover, the average office area shows that each lot in Midtown BD has on average, around 220,000 sq.~ft. of office area spread across different floors (e-bottom). In SDBD, however, the average office area per lot is around 73,000 sq.~ft. (d-bottom) which is significantly smaller in comparison with Midtown BD and FD. 
Hence, we can say that lot size can be one of the main factors in explaining why SDBD was not successful in absorbing the office demand. To accommodate such demand, there is a need for merging of the lots. Besides, the relatively small lot size is a perfect choice for private developers to place their next residential or mixed residential-commercial buildings and the upward trend of residential buildings shown in Figure~\ref{fig:case-study}(c-bottom), confirms this. By examining the map of office and residential buildings in SDBD, Figure~\ref{fig:interface}(b-bottom), we can see why there was no major increase in the number of offices in Downtown Brooklyn, as their lot sizes are limited and irregular shapes  complicates the merging possibility \citep{ellickson2012law}. Hence, either the city needs to look through mechanisms to incentivize zoning lot mergers or they have to look for other locations that satisfy the requirements such as access transit, locality and density. 

The analysis of the Downtown Brooklyn rezoning plan using \system{} enabled us to drill down one of the main reasons behind the detour of the plan from its initial goals, \emph{the lot size and shape}. This result would have not been reached without interactively exploring the data in fine-grain temporal intervals, tracking the changes on multiple spatial levels while transitioning between visualizing the change and values on the map to gain more insights. 
The results signify the need for regular reevaluation of the employed strategies, which can open the possibility of making early adjustments to ensure that the initial goals are realistic and attainable and the undertaken strategies are leading to the desired outcomes. As we showed, \system{} can provide a solid basis to perform such study. 

\begin{figure}[t!]
    \centering
    \includegraphics[scale = 0.13]{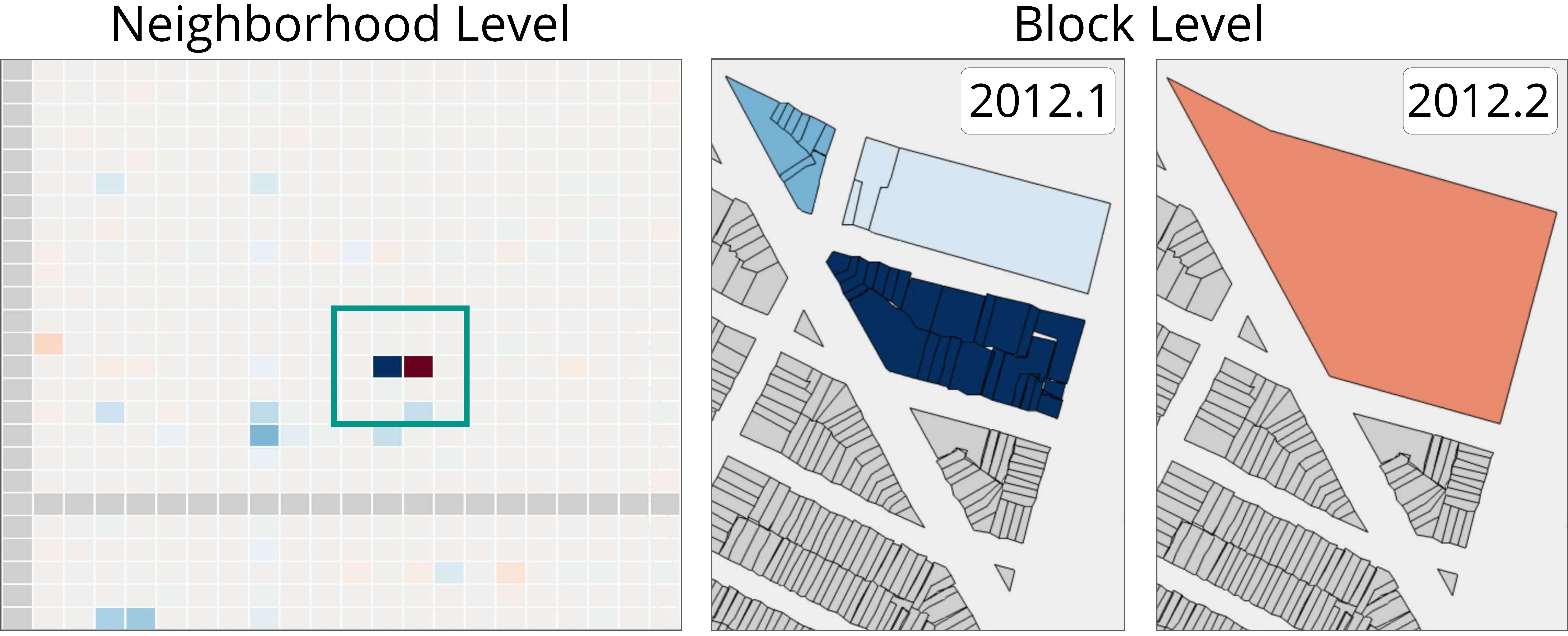}
    \caption{The heat matrix of Brooklyn's neighborhoods (left) shows an outlier (green square) in the variation of the number of \emph{open space and recreational} lots in Park Slope-Gowanus in 2012. Checking the block level, we see that the 2012.1 blocks were merged in 2012.2, during the construction of the Barclay's Center arena.}
    \label{fig:merge}
\end{figure}
\section{Discussion}
\label{sec:discussion}
\myparagraph{Lot merges and splits.}
As we observed in Section~\ref{sec:data-req}, lot merges and splits information is not provided in PLUTO data. 
Also, because of the inconsistencies in the lots' geometries and the reuse of the BBL of a split or merged lot after the operation, designing an algorithm to build the ``genealogy graph'' of the lots is a challenging task.
%
%
However, \system{} can facilitate the exploration of lot merges and splits throughout the city. Figure~\ref{fig:merge} depicts the merge of a group of lots detected using the heat matrix component of \system{}. In fact, the variation in the number of lots classified as open space and recreational land use in the Park Slope-Gowanus neighborhood using the heat matrix of the \emph{Data History}, shows a significant rise in 2012.1 (highlighted in dark blue) followed by a sharp drop in 2012.2 (highlighted in dark red). Drilling down to the blocks level, we observe that the pattern is being caused by the reclassification of the lots of three specific blocks, that were merged in the process of Barclay's Center construction in 2012.

\myparagraph{ArcGIS Pro comparison.}
ArcGIS Pro is a general, commercial GIS software designed for map making and spatial analysis. However, being a general system means it cannot be very efficient in handling all types of data, with higher than optimal execution time. For instance, the operations of selecting the Manhattan and Brooklyn boroughs, selecting a set of attributes, aggregating to the community district level and saving the results for three different MapPLUTO layers took more than 2 minutes to complete. 

Unlike \system, in ArcGIS Pro, the user cannot dynamically switch between different geographical levels of the map. They need to use aggregation functions for each layer to perform any type of analysis on that geographical level. The execution time can vary depending on the size of data and the level of aggregation. For instance, aggregating on block level for the three layers of MapPLUTO took slightly more than one minute whereas \system does this task interactively on-the-fly. Moreover, the process of measuring and visualizing the attribute and geometry change between two years where each year comes in a separate shapefile involves several steps in ArcGIS Pro such as Overlay Layers, Overlay Count, calculate year field and Heat Chart and this process should be repeated for any given pairs of years for which the changes should be computed. Doing the same task in \system{} is as easy as choosing the attribute of interest to see its change over a year or a selected interval of years. \\


\myparagraph{Experts feedback.}
%
During the development of \system{} we held semi-structured interviews with three urban planner partners, where we presented the different iterations of the system. During the interviews, we observed how easy it was for the participants to reach insights illustrated by our case studies and recorded their feedback about the usability of \system{}. It is important to point out that interviewees use the PLUTO data set as a part of their daily workflow, exploring various urban and environmental design, public policy, and planning scenarios. Our collaborators have extensive experience working for public agencies and large architectural firms (at least 6 years). Besides that, one of the coauthors of this paper is a specialist in urban systems and helped us with the assessment of the tool by using it to gain new insights into problems she is currently tackling in her research.

Users highlighted the usefulness and the value of the tool, especially its fast response time when querying PLUTO. One of the domain experts that used our tool is an employee at NYC's Department of City Planning, the agency responsible for maintaining the PLUTO data set. She highlighted that \system{} \emph{``merges PLUTO in a way that it's difficult to do, making visual exploration accessible beyond conventional tools"}.
Common suggestions include the need for a method to automatically load new data sets as they are being released and proper documentation.
We are currently incorporating some of the suggestions made by the different domain experts so that the tool can be widely deployed to different stakeholders, including city agencies.
It is important to point out that the case studies were performed independently by the domain experts, without supervision from the visualization experts.
\section{Conclusion and Future Work}
In this paper, we proposed \system{}, a system designed to enable the interactive visual exploration of all versions of the PLUTO data set released from 2002 to 2017. The data set can be seen as a spatiotemporal one in which the spatial component is not merely composed of points but of complex geometries that evolve over time.
To load all PLUTO releases in memory, we designed a redundancy removal strategy that allows the reuse of 90\% of the lots geometries and from 18\% to 70\% of the attributes depending on their classification. Also, in order to achieve on-the-fly aggregation of attributes and filtering of lots at multiple resolutions, we developed a data storage and a query processor that allow the processing of all lots of Manhattan in less than one second with a memory~overhead~of~only~3.5$\times$.

We intend to extend the use of the system to other cities that have similar data sets. Although the ideas presented in this paper can be used with any spatiotemporal data set with complex geometries, other challenges can emerge based on the characteristics of the new data. 
Also, building a visualization system to interactively design and explore different rezoning plan scenarios can be an interesting future research topic. Having access to such a system, urban planners would try to avoid unforeseen scenarios like the one described in Section~\ref{sec:case-studies}. 
Finally, we plan to investigate other visualization designs for spatiotemporal data sets with complex geometries.

\section*{Acknowledgment}
%
This work was supported in part by: NSF awards CNS-1229185, CCF-1533564, CNS-1544753, CNS-1730396, CNS-1828576, CNS-1626098;
CNPq grant 305974/2018-1; FAPERJ grants E-26/202.915/2019, E-26/211.134/2019.



\printcredits
\clearpage
\bibliographystyle{model5-names}
\bibliography{paper}





\end{document}